\documentclass{article}
\input{tcilatex}

\begin{document}

\title{\textbf{ON PROBABILITIES OF E2 TRANSITIONS BETWEEN POSITIVE-PARITY
STATES IN }$^{160}$\textbf{Dy NUCLEUS}}
\author{J. Adam$^{1,2}$, V.P. Garistov$^{1,3}$, M. Honusek$^{2}$, J. Dobes$%
^{2}$, I. Zvolski$^{2}$, \and J. Mrazek$^{2}$, A.A. Solnyshkin$^{1}$ \\
%EndAName
$^{1}${\small Joint Institute for Nuclear Research, Dubna, Russia}\\
$^{2}${\small Institute of Nuclear Physics, Czech Academy of Sciences, \v{R}e%
\v{z}}\\
$^{3}${\small Institute for Nuclear Research and Nuclear Energy, Sofia,
Bulgaria}}
\maketitle

\begin{abstract}
Reduced probabilities B(E2) of $\gamma $ transitions between states of
positive parity in the $^{160}$Dy nucleus are calculated within the
framework of the interacting boson model (IBM-1). The results are compared
with the experimental data.
\end{abstract}

Thorough and comprehensive investigations of the $^{160}Er\rightarrow
^{160m,g}Ho\rightarrow ^{160}Dy$ decay have yielded new extensive
experimental data on excited states and $\gamma $ transitions between them
in the $^{160}Dy$ nucleus [1]. In particular, observation of over 100 new
levels and over 500 new $\gamma $ transitions is reported and about 150
multipolarities are said to be established for the first time. The $%
^{160m,g}Ho\rightarrow ^{160}Dy$ decay scheme proposed by the authors
includes practically all known and newly discovered $\gamma $ transitions
except few of them whose total intensity is 0.9\% of decays. In [2] we made
an attempt to describe theoretically energy positions of most excited states
of positive parity and to reproduce energies of levels belonging to ground
rotational, gamma, and other rotational bands manifesting themselves in this
nucleus.

In this paper, using our data [1] on intensities and energies of $\gamma $
transitions de-exciting positive-parity levels with known $T_{1/2}$ [3], we
found experimental values of reduced probabilities $B(E2)$ by the formula%
\begin{equation}
B(E2)_{\gamma }=\frac{ln2}{T_{1/2}}\frac{I_{\gamma }10^{2}}{\sum I_{tot}%
\text{ }1.23E_{\gamma }^{5}}\ ,
\end{equation}

where $\sum I_{tot}$ is the sum of total intensities of al $\gamma $
transitions de-exciting a particular level. For $\gamma $ transitions
presumably exciting but not yet observed experimentally only limiting $B(E2)$
values were found from the upper bounds of their intensities experimentally
determined by us and the energies known from the positions of their
corresponding levels.

The thus obtained probabilities $B(E2)$ are compared with our calculations
within the interacting boson model $(IBM-1)$ determining the transition
operator $T(E2)$ as [4]: 
\begin{equation}
T(E2)=\alpha (d^{\dagger }s+s^{\dagger }\widetilde{d})^{(2)}+\frac{\beta }{%
\sqrt{5}}(d^{\dagger }\widetilde{d})^{(2)}\ .
\end{equation}
The absolute values $B(E2,2_{2}^{+}\rightarrow 0_{1}^{+})=4.43(34)$ and $%
B(E2,2_{1}^{+}\rightarrow 0_{^{_{1}}}^{+})=192.6(88)[W.u.]$ measured by us
were used in the calculations to determine the parameters $\alpha =1.914$
and $\beta =1.515$ .

The results of comparing experimental and calculated $B(E2)$ for each of the
levels with the known $T_{1/2}$ are presented in Tables 1.0--1.6, where
columns 2 show characteristics of the levels under consideration, including
the sum of intensities of their complete de-excitation, and columns 3, 4, 5
show characteristics of particular $\gamma $ transitions connecting them
with the corresponding final states whose characteristics are given in
columns 6 and 7. It is evident from Tables 1.0--1.6 (see columns 8 and 9)
that calculations are in a good agreement with the experimental values.

In Tables 2.0--2.19 experimental $B(E2)$ are compared with theoretical
values for particular excited states whose half-live times $T_{1/2}$ are
unknown but which have one or a few $\gamma $ transitions with the known
multipolarity $E2$. Then experimental reduced probabilities $B(E2)$ of $%
\gamma $ transitions from these levels were calculated by using the data [1]
and the relation%
\begin{equation}
B(E2)_{\gamma }=B(E2)_{0}\frac{I_{\gamma }E_{0}^{5}}{I_{0}\text{ }E_{\gamma
}^{5}}\ ,
\end{equation}

Here the lower index $``o\textquotedblright $\ marks the theoretical value
of $B(E2)_{0}$ calculated for each level using the $IBM-1.$ Further, taking
this theoretical value of $B(E2)_{0}$ as experimental value and using (3)
one can estimate other competing $\gamma $ transitions for the
experimentally known energies $E_{\gamma }$ and intensities $I_{\gamma }$\
of the corresponding $\gamma $ transition of $E2$ type. In the tables so
calculated values of $B(E2)$\ are marked with the symbol $``\equiv
\textquotedblright $. There is rather good agreement between theory and
experiment for a majority of $\gamma $ transitions with the known transition
type and multipolarity. The $\gamma $ transitions with considerable
disagreement between theoretical and experimental $B(E2)$ values make it
only possible to conclude that these transitions are not of $E2$ type, while
those with only slight disagreement between these values allow an assumption
that these transitions might be of $E2$ type. However, it is only after the
experiment that the final conclusions can be drawn.

$\bigskip $The investigation was supported by the RFBR.

\end{document}